\RequirePackage{fix-cm}

\documentclass[smallcondensed,natbib,referee]{svjour3}     

\smartqed  
\usepackage[pdftex]{graphicx}
 \usepackage{mathptmx}      

\usepackage[T1]{fontenc}
\usepackage[latin1]{inputenc}
\usepackage{amsfonts,amssymb,amsmath}

\usepackage{hyperref}
\usepackage[english]{babel}

\usepackage{color}

\journalname{Annals of Operations Research}

\begin{document}

\title{On benefits of cooperation under strategic power}
\author{M. G. Fiestras-Janeiro \and 
	I.	Garc\'{\i}a-Jurado \and 
	A. Meca \and
	M. A. Mosquera
}


\institute{M. G. Fiestras-Janeiro \at
	Departamento de Estat\'{\i}stica e Investigaci\'on Operativa. Universidade de Vigo, Spain. \\
	\email{fiestras@uvigo.es}           
	\and
	I.	Garc\'{\i}a-Jurado \at
	Departamento de Matem\'{a}ticas. Universidade da Coru\~{n}a, Spain.\\
	\email{ignacio.garcia.jurado@udc.es}
	\and
	A. Meca \at
	I. U. Centro de Investigaci\'on Operativa. Universidad Miguel Hern\'andez de Elche, Spain.\\
	\email{ana.meca@umh.es}
	\and
	M. A. Mosquera \at
	Corresponding author.\\
	Departamento de Estat\'{\i}stica e Investigaci\'on Operativa. Universidade de Vigo, Spain. \\
	\email{mamrguez@uvigo.es} 
}

\date{Received: date / Accepted: date}
\maketitle

\begin{abstract}
We introduce a new model involving TU-games and exogenous structures. 
Specifically, we consider that each player in a population can choose an element in a strategy set and that, for every possible strategy profile, a TU-game is associated with the population. This is what we call a {\em  TU-game with strategies}. We propose and characterize the maxmin procedure to map every game with strategies to a TU-game. We also study whether or not the relevant properties of TU-games are transmitted by applying the maxmin procedure. Finally, we examine two relevant classes of TU-games with strategies: airport and simple games with strategies.

\end{abstract}

\section{Introduction}
After its introduction in \cite{vonneumann44}, the model of TU-games has been widely used to analyze cooperation in multi-agent decision problems. Just to give few examples, \cite{Cetiner13} applied TU-games in revenue sharing for airline alliances, \cite{Fiestras15} in inventory control, \cite{Lozano15} in production processes, \cite{Naber15} in allocating CO$_2$ emission,  \cite{Kimms16} in logistics, \cite{Li16} in efficiency evaluation, \cite{Goyal17} in traffic management, and \cite{Balog17} in finance. In the seventies, several authors dealt with cooperative situations in which the cooperation was somewhat restricted by exogenous structures. For instance, \cite{Aumann74} treated TU-games with a coalition structure, \cite{Myerson77} studied TU-games with graph-restricted communication, and \cite{Owen77} considered TU-games with a priori unions. Since then, a number of models and solutions for TU-games with structures have been studied in the literature. A couple of recent papers on this topic are \cite{Alonso15} and \cite{Fernandez16}. \cite{Bilbao00} is a book devoted to this topic.

In this paper, we introduce a novel model involving TU-games and exogenous structures. Specifically, we consider that each player in a population chooses one of the strategies in a strategy set. Next, a TU-game that depends on the selected strategy profile is associated with the population.  This is what we call a {\em TU-game with strategies}. This model can be useful in a number of practical situations. For instance, consider a set of agents that have to divide an amount of money; they can negotiate directly or, on the contrary, they can previously take some (costly) actions that will modify their negotiation power. Any situation of this type can be modeled as a TU-game with strategies. A reasonable recommendation for the players involved in one such process is that they negotiate directly, avoiding the costly actions, but taking into account their capacities for changing the negotiation power. This is the main idea of our approach to TU-games with strategies: to associate each TU-game with strategies to a new TU-game that appropriately reflects the bargaining coalitional power of the involved players. 

As far as we know, this approach is not used in the literature to analyze situations involving strategic and cooperative environments. Instead, the existing literature tried to reduce the initial model to a strategic game in order to study its values and solutions. For instance, \cite{Ui00} introduced the model of TU-games with action choices. This mathematical model is similar to our model but the value of a coalition in any TU-game is not affected by the strategies of players outside that coalition. Moreover, its analysis focused on reducing a TU-game with action choices to a strategic game where the payoff of each strategy profile is determined by the Shapley value of the corresponding TU-game. In the same line, \cite{Brandenburger07} introduced the class of biform games where the value of each coalition depends on the strategies of all players, like in our model. Nevertheless, their analysis is similar to that in \cite{Ui00}, in the sense that they reduce a biform game to a strategic game, where the payoff of each strategy profile is determined by a particular element in the core of the corresponding TU-game. \cite{Platkowski16} introduced the class of evolutionary coalitional games. Again, the initial situation is transformed into a strategic game where the payoffs are obtaining by redistributing the sum of the initial individual payoffs using a well-known solution concept for TU-games.

Let us see now a few examples that fit our model and can be used to motivate our approach. Consider a situation in which three heirs, at first symmetrical, share an inheritance of three million euros and want to divide it. The heirs 1 and 2 can take legal actions that cost $0.25$ million euros for each of the three. If heirs 1 and 2 take these actions, they will keep the entire inheritance. However, heir 3 can respond with new legal actions; if heir 3 responds, all heirs will have an additional cost of $0.25$ million euros each and the final result will be that heirs 1 and 2 keep two millions and the remaining  million is divided among the three. This situation can be modeled as the following TU-game with strategies. 

\begin{center}
	\begin{tabular}{|c|c|c|}
		\hline
		NR & NL & L \\ \hline
		NL & u  & u \\ \hline
		L  & u  & v \\ \hline
	\end{tabular}
	\hspace*{0.5cm}
	\begin{tabular}{|c|c|c|}
		\hline
		R  & NL & L \\ \hline
		NL & u  & u \\ \hline
		L  & u  & w \\ \hline
	\end{tabular}
\end{center}
In this game heir 1 chooses row, heir 2 chooses column, and heir 3 chooses box. NL and L stand for "no legal actions" and "legal actions", respectively; NR and R stand for "no react" and "react", respectively. Finally, $N=\{ 1,2,3\}$ and $u,v,w$ are TU-games with set of players $N$ given by:
\begin{itemize}
	\item $u(S)=0$ for all $S\subsetneq N$, $u(N)=3$.
	\item $v(1)=v(2)=v(3)=-0.25$, $v(12)=2.5$, $v(13)=v(23)=-0.5$, $v(N)=2.25$.
	\item $w(1)=w(2)=w(3)=-0.5$, $w(12)=1$, $w(13)=v(23)=-1$, $w(N)=1.5$.
\end{itemize}
A sensible advise for the heirs can be the following: do not litigate, negotiate, but take into account in your negotiation the litigation options that you have. This advise is incorporated in our approach to TU-games with strategies because we build a new TU-game that takes into account the litigation power of the involved players and, then, we suggest that they use such a new game to allocate the inheritance. Since the main feature here is that heirs 1 and 2 can guarantee one million, the TU-game that seems to reflect the real power of the heirs is $\bar{u}$ given by $\bar{u}(12)=1$, $\bar{u}(S)=u(S)$ for all other $S\subset N$. Using the Shapley value,\footnote{For an introduction to cooperative games and to the Shapley value, \cite{Gonzalez10} can be consulted.} $\Phi$, to allocate the inheritance, the results is $\Phi(\bar{u})=(7/6,7/6,4/6)$ which takes into account the negotiation power of heirs 1 and 2.

Consider now a cost allocation situation with three companies, $N=\{ 1,2,3\}$,  involved in the realization of three projects. Each of these projects is an extension of the previous and has a cost of 100, 200, or 300 millions of euros. This means, for example, that project 2, which costs 200 millions, is an extension of project 1, with a cost of 100 millions. There is a first stage in which company $1$ can choose between applying for a subsidy ($a$), or not applying for a subsidy ($b$), for the realization of all the projects. If such a subsidy is granted, the cost of each project is reduced by 10 millions of euros.  Depending on the choice of company $1$, the resulting cost allocation problems (in millions of euros) are respectively described by the following airport games (\citealp{Littlechild73}):
\begin{itemize}
\item
$c_a(1)=90$, $c_a(2)=190$, $c_a(3)=290$, $c_a(12)=190$, $c_a(13)=c_a(23)=c_a(N)=290$,
\item
$c_b(1)=100$, $c_b(2)=200$, $c_b(3)=300$, $c_b(12)=200$, $c_b(13)=c_b(23)=c_b(N)=300$.
\end{itemize}
Assume that, the three companies decide to cooperate and to allocate the costs using the Shapley value, $\Phi$. It seems reasonable to conclude that company $1$ will choose to apply for the subsidy ($a$) and the cost allocation problem will be $c_a$. However, the Shapley value of $c_a$ is $\Phi (c_a)=(30,80,180)$ and it does not take into account the special contribution of company $1$.
We propose the following alternative approach. A new cost game ${\bar c}$ is built; it associates with every coalition $S$ the minimum cost that $S$ can guarantee making the right choices under its control. So,
$${\bar c}(S)=\left\{
\begin{array}{ll}
c_a(S)&\mbox{ if } 1\in S\\
c_b(S)&\mbox{ if } 1\not\in S,\\
\end{array}
\right.$$
i.e., ${\bar c}(1)=90$, ${\bar c}(2)=200$, ${\bar c}(3)=300$, ${\bar c}(12)=190$, ${\bar c}(23)=300$, ${\bar c}(13)={\bar c}(N)=290$. Observe that $\Phi ({\bar c})=(23.33,83.33,183.33)$ seems to be a more reasonable allocation of the total cost in this situation, due to the special contribution of company $1$. Indeed, the Shapley value for this alternative situation divides equally the subsidy of 10 millions, i.e. 10/3, and adds this amount to the previous costs of companies 2 and 3,  while reducing in 20/3 the cost of company 1. 

Finally, consider a Parliament with $33$ seats. Three political parties are represented in the Parliament, each of them having $11$ seats. Because of political reasons, parties $1$ and $3$ are unable to reach stable agreements between them. Therefore, party $2$ must choose between reaching a stable agreement with party $1$ or party $3$.  A stable agreement between two parties means that they will negotiate a common position for every initiative to be voted in the Parliament. In this situation, party $2$ will choose its partner; whether party $1$ or party $3$ are chosen, the power of the coalitions are respectively described by the following simple games $v_1$ and $v_3$:
\begin{itemize}
\item
$v_1(1)=v_1(2)=v_1(3)=v_1(13)=v_1(23)=0$, $v_1(12)=v_1(N)=1$.
\item
$v_3(1)=v_3(2)=v_3(3)=v_3(12)=v_3(13)=0$, $v_3(23)=v_3(N)=1$.
\end{itemize}
In order to measure the power of each party in this situation before a stable agreement is made, we  propose the following approach. Like in the example above, a new simple game ${\bar v}$ is built; again, it associates with every coalition $S$ the power that $S$ can guarantee making the right choices under its control. In this case:
$${\bar v}(S)=\left\{
\begin{array}{cl}
\max\{v_1(S), v_3(S)\}&\mbox{ if } 2\in S\\
0&\mbox{ if } 2\not\in S,\\
\end{array}
\right.$$
i.e., ${\bar v}(1)={\bar v}(2)={\bar v}(3)={\bar v}(13)=0$, ${\bar v}(12)={\bar v}(23)={\bar v}(N)=1$. Using the Shapley value, $\Phi$, as a power index, we obtain $\Phi ({\bar v})=(1/6,4/6,1/6)$ which seems to be a reasonable measure of the power in this situation.

These three examples illustrate situations that can be modeled as TU-games with strategies. In each of these games, there is a set of players, a set of strategies for every player, and a map that associates a TU-game with every possible strategy profile. In order to allocate the benefits arising from cooperation to the players, we build a new TU-game that maps every possible coalition to the optimal value that it can guarantee based on the strategies selected by its members. This procedure to build a TU-game is also used in other environments like strategic games. For instance, \cite{Carpente05} analyzed strategic games where players can cooperate and used this procedure to build a TU-game from the strategic game in order to define a value for it. Differently, \cite{Kalai13} used cooperation to define values for strategic games by decomposing each strategic game to cooperative and competitive components.

In the next section, we formally introduce and analyze the procedure used to map a TU-game with strategies to a TU-game. We also provide an axiomatic characterization of this procedure. In Section 3, we determine which properties of the TU-games defining a TU-game with strategies are transmitted through the procedure. In Section 4, we analyze the applications of TU-games with strategies for airport games and simple games. Finally, Section 5 presents our conclusions.

\section{TU-games with strategies}

Fix $\cal{N}$, a universe of potential players. Let $N\subset \cal{N}$ be a finite set of players.  A transferable utility game (in short a TU-game) is a function $v:2^N\longrightarrow \mathbb{R}$ such that $v(\emptyset)=0$. We denote by $G(N)$ the set of TU-games with a finite player set $N$ and by $G$ the set $\cup_{N\subset \cal{N}}G(N)$.
\begin{definition}
A TU-game with strategies with player set $N$ is a pair $(X,V)$ such that:
\begin{itemize}
\item
$X=\prod_{i\in N}X_i$ is a finite set of strategy profiles, being $X_i$ the player's $i$ strategy set for every $i\in N$, and
\item
$V:X\rightarrow G(N)$ is a map that associates a TU-game $V(x)$ with every $x\in X$.
\end{itemize}
We denote by $SG(N)$ the set of TU-games with strategies and player set $N$ and by $SG$ the set $\cup_{N\subset \cal{N}}SG(N)$.
\end{definition}

Notice that a TU-game with strategies is the same mathematical model as a biform game, introduced in \cite{Brandenburger07}. However, the interpretation of the model is completely different in this paper. Here, we assume that players can coordinate their strategies and cooperate, and that the grand coalition aims to allocate among its members the benefits of the cooperation by considering their strategic power. So, we focus on a procedure to transform a TU-game with strategies into a TU-game. We do it using an axiomatic approach, i.e. we first identify some properties that are appropriate for such a procedure and then we prove that there exists only one procedure satisfying those properties. The properties are inspired by analogous ones introduced in \cite{Carpente05}.

\begin{definition}
	A procedure to transform a TU-game with strategies into a TU-game (in brief a {\em procedure}) is a map $\phi:SG\rightarrow G$ that associates a TU-game $\phi(X,V)\in G(N)$ with every TU-game with strategies $(X,V)\in SG(N)$.
\end{definition}

Let us state some properties that are appropriate for a procedure $\phi$.

\begin{description}
	\item[Individual objectivity.] For every $(X,V)\in SG(N)$, if a player $i\in N$ is such that $V(x)(i)=c$, for all $x\in X$, then $\phi (X,V)(i)=c.$
\end{description}
Individual objectivity states that if the coalition consisting of player $i$ receives the same utility $c$ for every possible strategy profile, the TU-game resulting from the procedure associates $c$ to coalition $\{i\}$.

\begin{description}
	\item[Monotonicity.] If $(X,V), (X,\bar{V})\in SG(N)$ satisfy that $V(x)\geq \bar{V}(x)$ for all $x\in X$,\footnote{Given $v,w\in G(N)$ we write $v\geq w$ if and only if $v(S)\geq w(S)$, for every $S\subset N$.} then $\phi (X,V)\geq \phi (X,\bar{V})$.
\end{description}	
Monotonicity states that if a game with strategies $(X,V)\in SG(N)$ associates to every coalition and strategy profile a utility greater than or equal to the corresponding utility associated by a different $(X,\bar{V})\in SG(N)$, then the TU-game for $(X,V)$ resulting from the procedure is greater than or equal to the TU-game for $(X,\bar{V})$.

Take $(X,V)\in SG(N)$, $i\in N$, and $S\subset N$ with $i\in S$. A strategy $x_i\in X_i$ of player $i$ is {\it weakly dominated} in $S$ if there exists a strategy $x_i'\in X_i$, $x'_i\neq x_i$, such that $V(\bar{x}_{-i},x'_i)(S)\geq V(\bar{x}_{-i},x_i)(S)$ for all $\bar{x}_{-i}\in \prod_{j\in N\setminus \{i\}}X_j$. Moreover, $(X^{-x_i},V)$ denotes the TU-game with strategies that is obtained from $(X,V)$ by deleting strategy $x_i$.
\begin{description}
	\item[Irrelevance of weakly dominated strategies.]	For any $(X,V)\in SG(N)$, $i\in N$, and $S\subset N$ with $i\in S$, if strategy $x_i\in X_i$ is weakly dominated in $S$, then $\phi (X,V)(S)=\phi (X^{-x_i},V)(S)$.
\end{description}	
Irrelevance of weakly dominated strategies states that if a player loses the ability to use a weakly dominated strategy (for a coalition $S$) in the original game with strategies, this should not affect the final result for $S$.

Take $(X,V)\in SG(N)$, $i\in N$, $j\in N\setminus\{ i\}$, and $S\subset N\setminus\{ j\}$ with $i\in S$. A strategy $x_j\in X_j$ of  player $j$ is a {\it weakly dominated threat} to player $i$ in $S$ if for every $\bar{x}_{-j}\in \prod_{k\in N\setminus \{j\} }X_k$ there exists a strategy $x'_j\in X_j$, $x'_j\neq x_j$, such that $V(\bar{x}_{-j},x'_j)(S)\leq V(\bar{x}_{-j},x_j)(S)$. In words, a strategy of player $j$ is a weakly dominated threat to player $i$ in $S$ if there exists a different strategy of player $j$ that can further harm player $i$ in $S$.
\begin{description}	
	\item[Irrelevance of weakly dominated threats.]	For any $(X,V)\in SG(N)$, players $i,j\in N$, $i\neq j$, and $S\subset N\setminus\{ j\}$ with $i\in S$, if strategy $x_j\in X_j$ is a weakly dominated threat to player $i$ in $S$,	then $\phi (X,V)(S)=\phi (X^{-x_j},V)(S)$.
\end{description}	
Irrelevance of weakly dominated threats states that if a player $j$ loses the ability to use a weakly dominated threat to other player $i$ (for a coalition $S$) in the original game with strategies, this should not affect the final result for $S$.

Take $(X,V)\in SG(N)$ and $\emptyset\neq S\subset N$. Denote by $N^S$ the set $\{[S]\}\cup N\setminus S$, i.e. the set of $|N|-|S|+1$ players in which the coalition $S$ is considered as a single player, and by $(X^S,V^S)$ the TU-game with strategies that is obtained from $(X,V)$ by considering the coalition $S$ 	as a single player.\footnote{$(X^S,V^S)$ is given by $X^S=X_{[S]}\times (\prod_{i\in N\setminus S}X_i)$ with $X_{[S]}=\prod_{i\in S}X_i$, and  for every $x^S\in X^S$ and every $T\subset N\setminus S$,  $V^S(x^S)(T)=V(x)(T)$, and $V^S(x^S)(T\cup[S])=V(x)(T\cup S)$.}
	
\begin{description}	
	\item[Merge invariance.] Take $(X,V)\in SG(N)$ and $\emptyset\neq S\subset N$, then, for all $T\subset N\setminus S$, $\phi (X,V)(T)=\phi (X^S,V^S)(T)$ and $\phi (X,V)(T\cup S)=\phi (X^S,V^S)(T\cup \{[S]\})$.
\end{description}	
Merge invariance states that a coalition of players cannot influence its worth by merging and acting as one player.

\begin{definition}
	The {\em maxmin procedure} is the map $\psi:SG\rightarrow G$ given, for all $N \subset \cal{N}$ and $(X,V)\in SG(N)$, by:
	$$\psi (X,V)(S)=\max_{x_S\in\prod_{i\in S}X_i}\min_{x_{N\setminus S}\in\prod_{i\in N\setminus S}X_i}V(x_S,x_{N\setminus S})(S)$$
	for all $S\subset N$.
\end{definition}

The next result provides an axiomatic characterization of $\psi$. It is an extension\footnote{ It is an extension because every finite strategic game $(\{X_i\}_{i\in N},\{H_i\}_{i\in N})$ can be seen as the game with strategies $(X,V)$ with $X=\prod_{i\in N}X_i$ and $V(x)(S)=\sum_{i\in S}H_i(x)$, for all $x\in X$ and all $S\subset N$.} of Theorem 7 in \cite{Carpente05} and its proof is similar to that of Theorem 7.  We include the proof in the Appendix for the sake of clarity.

\begin{theorem}\label{thm:proc}
	The maxmin procedure is the unique procedure to transform a TU-game with strategies into a TU-game that satisfies individual objectivity, monotonicity, irrelevance of weakly dominated strategies, irrelevance of weakly dominated threats and merge invariance.
\end{theorem}

Notice that this procedure is the one that we used in the introduction of this paper to tackle the three examples presented there. In order to verify the procedure, we will now revisit the same three examples.

\begin{example}
	We consider the case of the three heirs willing to divide the inheritance.
	That situation can be modeled as a TU-game with strategies $(X,V)$ such that:
	\begin{itemize}
		\item
		$X=X_1\times X_2\times X_3$, where $X_1=X_2=\{ NL,L\}$, $ X_3=\{ NR,R\}$.
		\item
		$V$ is given by $V(L,L,NR)=v$, $V(L,L,R)=w$ and $V(x)=u$ for all other $x\in X$.
	\end{itemize}
	It is easy to check that
	$\psi (X,V)(S)={\bar u}(S)$ for all $S\subset N$.
\end{example}

\begin{example}
	We now consider the case of the three companies involved in the realization of three projects whose costs depend on the company 1's choice between applying for a subsidy (a) or not applying for a subsidy (b). Recall that the cost  functions are given by 
	\begin{itemize}
		\item
		$c_a(1)=90$, $c_a(2)=190$, $c_a(3)=290$, $c_a(12)=190$, $c_a(13)=c_a(23)=c_a(N)=290$,
		\item
		$c_b(1)=100$, $c_b(2)=200$, $c_b(3)=300$, $c_b(12)=200$, $c_b(13)=c_b(23)=c_b(N)=300$.
	\end{itemize}

	This situation can be modeled as the TU-game with strategies $(X,C)$ such that:
	\begin{itemize}
		\item
		$X=X_1\times X_2\times X_3$, where $X_1=\{ a,b\}$, $X_2=\{ \alpha\}$, $ X_3=\{ \beta\}$.
		\item
		$C$ is given by $C(a, \alpha, \beta)=c_a$ and $C(b, \alpha, \beta)=c_b$.
	\end{itemize}
	Notice that $c_a$ and $c_b$ are cost TU-games. Then, for preserving the philosophy of the maxmin procedure we have to interchange maximum and minimum in its definition; with this in mind, it is easy to check that $\psi (X,C)(S)={\bar c}(S)$, for all $S\subset N$.
\end{example}

\begin{example}
	Finally we consider the Parliament described in the Introduction. This situation can be modeled as the TU-game with strategies $(X,V)$ such that:
	\begin{itemize}
		\item
		$X=X_1\times X_2\times X_3$, where $X_2=\{ 1,2\}$, $X_1=\{ \alpha\}$, $ X_3=\{ \beta\}$.
		\item
		$V$ is given by $V( \alpha, 1,\beta)=v_1$ and $V( \alpha, 2,\beta)=v_2$.
	\end{itemize}
	Then, it is easy to check that $\psi (X,V)(S)={\bar v}(S)$, for all $S\subset N$.
\end{example}
\section{Inheritance of properties}
In this section we analyze which  properties are transmitted by the maxmin procedure. We want to know if for every TU-game with strategies $(X,V)$ such that  the TU-game $V(x)$ satisfies some property   for every strategy profile $x$, it is true that the TU-game $\psi(X,V)$ also satisfies this property. In particular, we study the properties of superadditivity, monotonicity, and balancedness. We recall the definition of these properties in the context of TU-games. Let $v\in G(N)$.
We say that $v$ is \textit{superadditive} if $v(S\cup T)\geq v(S)+v(T)$ for every $S,T\subset N$ with $S\cap T=\emptyset$.  We  say that $v$ is \textit{monotone} if $v(S)\leq v(T)$ whenever $S\subset T$.  We say that $v$ is \textit{balanced} if $Core(v)\neq \emptyset$ being 
\begin{equation*}
Core(v)=\{x\in \mathbb{R}^N:\ \sum_{i\in S}x_i=v(N)\text{ and }\sum_{i\in S}x_i\geq v(S)\text{ for every }S\subset N\}.
\end{equation*}
Next theorem shows that the maxmin procedure transmits the superadditivity property.
\begin{theorem}
Let $(X,V)\in SG(N)$. If $V(x)$ is superadditive for all $x\in X$, then $\psi(X,V)$ is also superadditive.
\end{theorem}

\begin{proof} 
Assume that $V(x)$ is superadditive for all $x\in X$. Consider the TU-game obtained by the maxmin procedure: 
\[\psi(X,V)(S)=\max_{x_S\in\prod_{i\in S}X_i}\min_{x_{N\setminus S}\in\prod_{i\in N\setminus S}X_i}V(x_S,x_{N\setminus S})(S),\]
for every $S\subset N$. 

Assume that $\psi(X,V)$ is not superadditive. Then, there exist two non-empty coalitions $S,T \subset N$ with $S\cap T = \emptyset$ such that 
\[\psi(X,V)(S\cup T)<\psi(X,V)(S)+\psi(X,V)(T).\]

For each $S\subset N$, let $x^S\in X$ be such that $\psi(X,V)(S)=V(x^S)(S)$. Take $\bar{x}_{N\setminus (S\cup T)} \in X_{N\setminus (S\cup T)}$ such that 
\[V(x^S_S,x^T_T,\bar{x}_{N\setminus (S\cup T)})(S\cup T) =\min_{x_{N\setminus (S\cup T)} \in \Pi_{i\in N\setminus (S\cup T) }X_i} V(x^S_S,x^T_T,x_{N\setminus (S\cup T)})(S\cup T).\]

Then,
\begin{eqnarray*}
\psi(X,V)(S\cup T) 
					 & \boldsymbol{<} & \psi(X,V)(S)+\psi(X,V)(T)\\ 
					 & \leq & V(x^S_S,x^T_T,\bar{x}_{N\setminus (S\cup T)})(S) + V(x^S_S,x^T_T,\bar{x}_{N\setminus (S\cup T)})(T)\\
  				 & \leq & V(x^S_S,x^T_T,\bar{x}_{N\setminus (S\cup T)})(S\cup T) \\
					 & = &  \min_{x_{N\setminus (S\cup T)} \in X_{N\setminus (S\cup T)}} V(x^S_S,x^T_T,x_{N\setminus (S\cup T)})(S\cup T)
\end{eqnarray*}
where the first inequality follows by assumption, the second one follows by definition of $\psi(X,V)$ and the choice of $x^S$ and $x^T$, the third one follows by the superadditivity of each $V(x)$, and the last one follows by definition of $\bar{x}_{N\setminus (S\cup T)}$.

But, this is a contradiction since 
\[\psi(X,V)(S\cup T)=\max_{x_{S\cup T}\in X_{S\cup T}}\min_{x_{N\setminus (S\cup T)}\in X_{N\setminus (S\cup T)}}V(x_{S\cup T},x_{N\setminus (S\cup T)})(S).\]
\qed
\end{proof}

Next result shows that the maxmin procedure transmits the monotonicity property.
\begin{theorem}\label{thm:mon}
	Let $(X,V)\in SG(N)$. If $V(x)$ is monotone for all $x\in X$, then $\psi(X,V)$ is also monotone.
\end{theorem}
\begin{proof}
	Assume that $V(x)$ is monotone for all $x\in X$. Consider the TU-game obtained by the maxmin procedure: 
	\[\psi(X,V)(S)=\max_{x_S\in\prod_{i\in S}X_i}\min_{x_{N\setminus S}\in\prod_{i\in N\setminus S}X_i}V(x_S,x_{N\setminus S})(S),\]
	for every $S\subset N$.  Let $S\subset T\subset N$. Then,
	\begin{eqnarray*}
	\psi(X,V)(S)&=&\max_{x_S\in\prod_{i\in S}X_i}\min_{x_{N\setminus S}\in\prod_{i\in N\setminus S}X_i}V(x_S,x_{N\setminus S})(S),\\
	&=&\max_{x_S\in\prod_{i\in S}X_i}\min_{x_{T\setminus S}\in\prod_{i\in T\setminus S}X_i}\min_{x_{N\setminus T}\in\prod_{i\in N\setminus T}X_i}V(x_S,x_{T\setminus S},x_{N\setminus T})(S)\\
	&\leq& \max_{x_S\in\prod_{i\in S}X_i}\max_{x_{T\setminus S}\in\prod_{i\in T\setminus S}X_i}\min_{x_{N\setminus T}\in\prod_{i\in N\setminus T}X_i}V(x_S,x_{T\setminus S},x_{N\setminus T})(T)\\
	&=&\psi(X,V)(T).
	\end{eqnarray*} \qed
\end{proof}
In general, the maxmin procedure does not transmit the balanced property as we see next.
\begin{example}\label{ex:coreempty}
	Let us take $N=\{1,2,3\},\quad X_1=\{U\}\quad X_2=\{L,R\}\quad X_3=\{F\}$. The TU-games associated with each strategy profile are:
	\[\begin{array}{|r|ccccccc|}
		\hline
		         S & \{1\} & \{2\} & \{3\} & \{1,2\} & \{1,3\} & \{2,3\} & \{1,2,3\} \\ \hline
		{V(U,L,F)} &   1   &   3   &   1   &    7    &    6    &    1    &     9     \\
		{V(U,R,F)} &   1   &   1   &   1   &    2    &    7    &    9    &    10     \\ \hline
	\end{array}\]
	It is easy to check that both games have non-empty cores. Nevertheless, the maxmin procedure provides the following TU-game:
	\[\begin{array}{|r|ccccccc|}
		\hline
		        S & \{1\} & \{2\} & \{3\} & \{1,2\} & \{1,3\} & \{2,3\} & \{1,2,3\} \\ \hline
		\psi(X,V) &   1   &   3   &   1   &    7    &    6    &    9    &    10     \\ \hline
	\end{array}
	\]
	Let us check now that $\psi(X,V)$ is not balanced. If the allocation $(x_1,x_2,x_3)$ belongs to the core of $\psi(X,V)$ we have $x_1=10-x_2-x_3$, $x_1\geq 1$, and $x_2+x_3\geq 9$. Then, $1\leq x_1\leq 1$. Since $x_1=1$, $x_1+x_2\geq 7$, and $x_1+x_3\geq 6$, we obtain $x_2\geq 6$ and $x_3\geq 5$. Thus, $x_1+x_2+x_3>10$ and we get a contradiction. Then, the core of the game $\psi(X,V)$ is empty.  
	\end{example}

We give next a result that may be helpful to study the core of $\psi(X,V)$ using information from the original TU-games with strategies.
\begin{theorem}\label{thm:core}
	Let $(X,V)\in SG(N)$. Then, $Core(\psi(X,V))=\cap_{x\in X}Core(V^{x})$ where for every $x\in X$, $V^{x}$ is the TU-game given by $V^{x}(N)=\psi(X,V)(N)$ and $V^{x}(S)=\min_{\bar{x}_{N\setminus S}}V(x_S,\bar{x}_{N\setminus S})(S)$, for every $\emptyset\neq S\subsetneq N$.
\end{theorem}
\begin{proof}
First, we check that  $Core(\psi(X,V))\subset\cap_{x\in X}Core(V^{x})$. Let  $a\in Core(\psi(X,V))$. By definition,
\begin{align*}
&\sum_{i\in S}a_i\geq \psi(X,V)(S),\text{ for every }\emptyset\neq S\subsetneq N, \\
&\sum_{i=1}^na_i=\psi(X,V)(N).
\end{align*}
Moreover, since $\psi(X,V)(N)=V^{x}(N)$ for all $x\in X$ and $\psi(X,V)(S)=\max_{x_S\in\prod_{i\in S}X_i}\min_{\bar{x}_{N\setminus S}\in\prod_{i\in N\setminus S}X_i}V(x_S,\bar{x}_{N\setminus S})(S)$ for every $\emptyset\neq S\subsetneq N$, it is hold, for every $x\in X$,
\begin{align*}
&\sum_{i\in S}a_i\geq \min_{\bar{x}_{N\setminus S}\in\prod_{i\in N\setminus S}X_i}V(x_S,\bar{x}_{N\setminus S})(S)=V^{x}(S), \text{ for every }\emptyset\neq S\subsetneq N,\\
&\sum_{i=1}^na_i=V^{x}(N)
\end{align*}
Then,  we obtain that $a\in Core(V^x)$ for every $x\in X$. Therefore,  $a\in \cap_{x\in X}Core(V^{x})$.

Second, we prove that  $\cap_{x\in X}Core(V^{x})\subset Core(\psi(X,V))$.  Let $a\in \cap_{x\in X}Core(V^{x})$. By definition, for every $x\in X$, 
\begin{align*}
&\sum_{i\in S}a_i\geq V^x(S) =\min_{\bar{x}_{N\setminus S}\in\prod_{i\in N\setminus S}X_i}V(x_S,\bar{x}_{N\setminus S})(S), \text{ for every }\emptyset\neq S\subsetneq N,\\
&\sum_{i=1}^na_i=V^{x}(N).
\end{align*}
Then,
\begin{align*}
&\sum_{i\in S}a_i\geq \max_{x\in X}V^x(S) =\psi(X,V)(S), \text{ for every }\emptyset\neq S\subsetneq N,\\
&\sum_{i=1}^na_i=\psi(X,V)(N)
\end{align*}
since $V^{x}(N)=\psi(X,V)(N)$ for every $x\in X$. Therefore, we obtain that  $a\in Core(\psi(X,V))$. \qed
\end{proof}
We revisit Example~\ref{ex:coreempty} to illustrate this result.
\begin{example}\label{ex:coreemptyc}
We have seen that  the core of $\psi(X,V)$ is empty in Example~\ref{ex:coreempty}. Here we analyze the core of each TU-game $V^x$ for every $x\in X$ to illustrate Theorem~\ref{thm:core}. Notice that the TU-games $V^{x}$ are given  by
\[\begin{array}{|r|ccccccc|}
\hline 
S & \{1\} & \{2\} & \{3\} & \{1,2\} & \{1,3\} & \{2,3\} & \{1,2,3\}\\
\hline
V^{(U,L,F)}& 1 & 3 & 1 & 7 & 6 & 1 & 10\\
V^{(U,R,F)}& 1 & 1 & 1 & 2 & 6 & 9 & 10\\ \hline
\end{array}
\]
In Figure~\ref{fig:exempty}  we depict\footnote{This figure has been built with the toolbox TUGlab of MATLAB$^{\circledR}$ (\citealp{Miras08}). The web page of TUGlab can be found in \url{http://eio.usc.es/pub/io/xogos/index.php}.}  the core of $V^{(U,L,F)}$ and $V^{(U,R,F)}$ described    as
	\begin{equation*}
	\begin{array}{l}
	Core(V^{(U,L,F)})=conv\{(4,3,3),(5,4,1),(3,4,3),(6,3,1)\},\\
	Core(V^{(U,R,F)})=conv\{(1,1,8),(1,4,5)\}.
	\end{array}
	\end{equation*}
It is clear that $Core(V^{(U,L,F)})\cap Core(V^{(U,R,F)})=\emptyset$.
\begin{figure}[!htb]
	\begin{center}
		\includegraphics[scale=0.45]{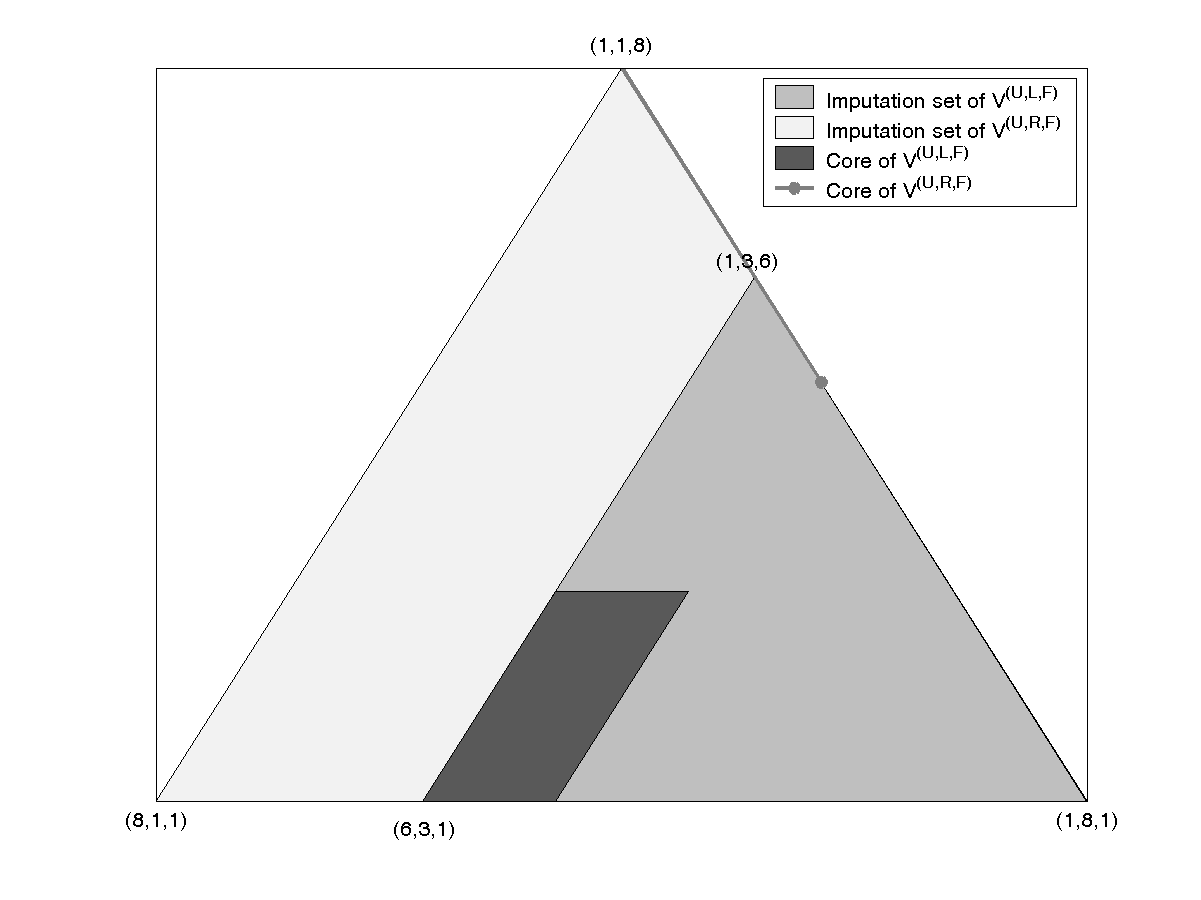} 
		\caption{$Core(V^{(U,L,F)})$ and $Core(V^{(U,R,F)})$ in Example~\ref{ex:coreemptyc}.}
		\label{fig:exempty}
	\end{center}
\end{figure}
\end{example}
A property related to balancedness of a TU-game is convexity since any convex game is balanced. Recall that a TU-game $v\in G(N)$ is \textit{convex} if and only if we have $v(T\cup i)-v(T)\geq v(S\cup i)-v(S)$, for every $i\in N$ and every $S,\ T\subset N\setminus \{i\}$ with $S\subset T$. The following counterexample shows that the maxmin procedure does not preserve the convexity property, in general.
\begin{example}\label{lab:exc}
Consider $N=\{1,2,3\},\quad X_1=\{U,D\}\quad X_2=\{L\}\quad X_3=\{F\}$. The TU-games associated with each strategy profile are:
\[\begin{array}{|r|ccccccc|}
	\hline
	         S & \{1\} & \{2\} & \{3\} & \{1,2\} & \{1,3\} & \{2,3\} & \{1,2,3\} \\ \hline
	{V(U,L,F)} &   2   &   1   &   3   &    4    &    7    &    4    &     9     \\
	{V(D,L,F)} &   1   &   4   &   2   &    5    &    3    &    6    &     9     \\ \hline
\end{array}\]
It is easy to check that both  TU-games are convex and then they have non-empty cores. The maxmin procedure provides the following TU-game:
\[\begin{array}{|r|ccccccc|} 
\hline
 S & \{1\} & \{2\} & \{3\} & \{1,2\} & \{1,3\} & \{2,3\} & \{1,2,3\}\\
\hline
\psi(X,V)& 2 & 1 & 2 & 5 & 7 & 4 & 9
	\\ \hline
\end{array}
\]
This game is not convex because 
$$\psi(X,V)(13)-\psi(X,V)(1)=5 > \psi(X,V)(123)-\psi(X,V)(12)=4$$
but, it has a non-empty core. For every $x\in X$, the TU-game $V^x$   is given by
\[\begin{array}{|r|ccccccc|}
	\hline
	          S & \{1\} & \{2\} & \{3\} & \{1,2\} & \{1,3\} & \{2,3\} & \{1,2,3\} \\ \hline
	V^{(U,L,F)} &   2   &   1   &   2   &    4    &    7    &    4    &     9     \\
	V^{(D,L,F)} &   1   &   1   &   2   &    5    &    3    &    4    &     9     \\ \hline
\end{array}\]
Notice that $(3,2,4)\in Core(V^{(U,L,F)})\cap Core(V^{(D,L,F)})$ and, thus $(3,2,4)\in Core(\psi(X,V))$. Figure~\ref{fig:ej1} shows the relationship among $Core(V^{(U,L,F)})$, $Core(V^{(D,L,F)})$, and $Core(\psi(X,V))$.
\begin{figure}[!htb]
\begin{center}
	\includegraphics[scale=0.45]{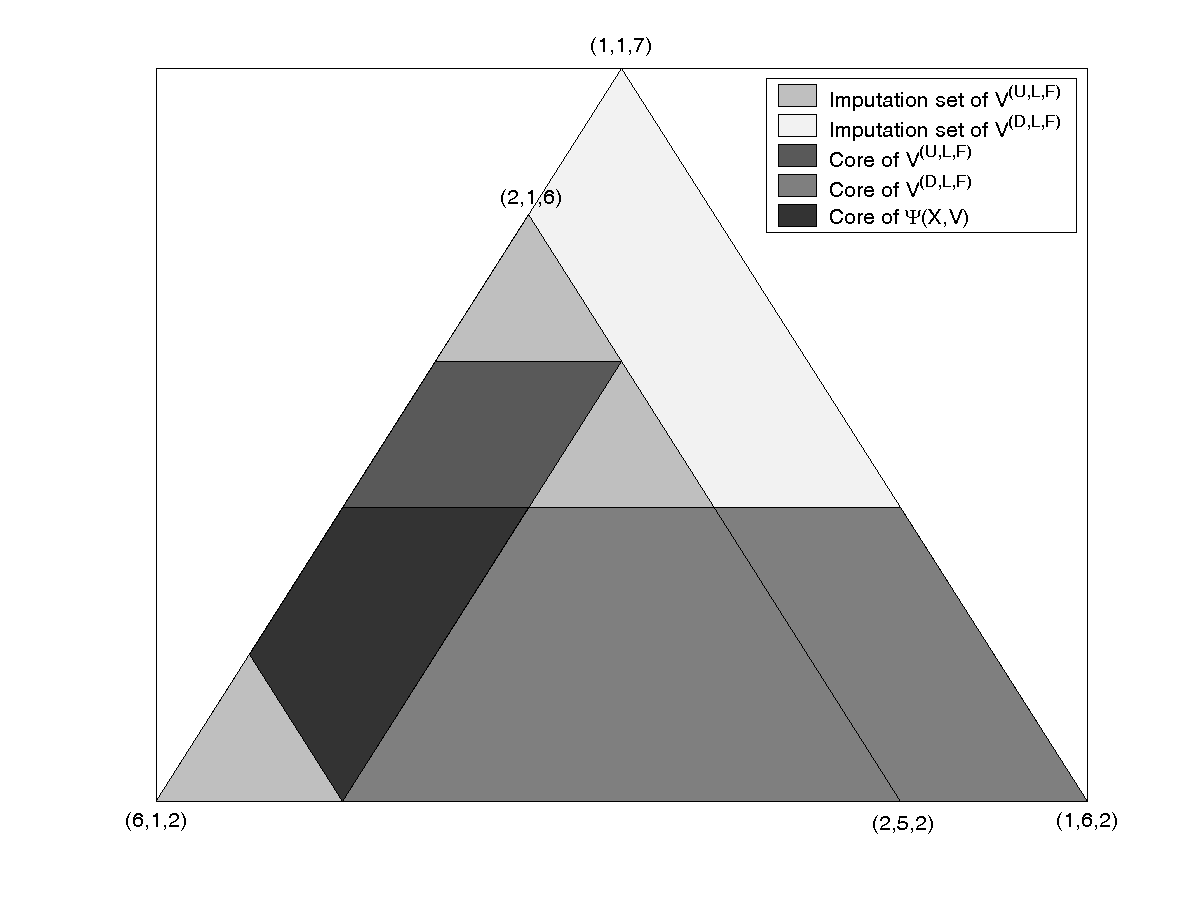}
\end{center}
\caption{$Core(V^{(U,L,F)})$, $Core(V^{(D,L,F)})$, and $Core(\psi(X,V))$ in Example ~\ref{lab:exc}. }\label{fig:ej1}
\end{figure}
\end{example}
\section{Two  particular cases}
In this section we study two specific cases of TU-games with strategies inspired by the examples described in the Introduction.  In Example 2, each TU-game associated to any strategy profile is an airport game.  In Example 3, each TU-game associated to any strategy profile belongs to the class of simple games. Airport games and simple games are well-known classes of TU-games which have many applications and have been widely studied in the literature.

\subsection{Airport games with strategies}
An \emph{airport problem} (\citealp{Littlechild73}) is described as follows. Suppose that ${\cal T}$ is the set of types of planes operating in an airport in a particular period. Denote by $N_{\tau}$ the set of movements that are made by planes of type $\tau\in {\cal T}$ and by $N$ the set of all movements, i.e. $N=\cup_{\tau\in {\cal T}}N_{\tau}$. Let $d_{\tau}$ be the cost of a runway that is suitable for planes of type $\tau$ in the considered period. Without loss of generality, we assume that 
$$0\leq d_1\leq d_2\leq\hdots\leq d_{|{\cal T}|}.$$ 
We can associate a cost game to this situation as follows.	For every $\emptyset\neq S\subset N$, $c(S)$ is defined as the cost of a runway that can be used by all the movements in $S$, i.e.
\begin{equation*}
c(S):=\max\{d_{\tau}: S\cap N_{\tau}\neq \emptyset\}
\end{equation*}
and $c(\emptyset)=0$. The TU-cost game $c$ is known as an \emph{airport game}.

An \emph{airport game with strategies} is a pair $(X,C)\in SG(N)$, where $C(x)$ is an airport game for all $x\in X$. It means that for all $x\in X$ and  $\emptyset\neq S\subset N$,
\begin{equation*}
C(x)(S)=\max\{C(x)(j):\ j\in S\}
\end{equation*}
given $C(x)(j)\geq 0$, for every $j\in N$.

Notice that we are dealing with cost TU-games for each strategy profile. Then, some definitions and properties established for TU-games has to change adequately. Namely, the \emph{core of a cost TU-game} $c$ is the core of the TU-game $-c$ and we say that a cost TU-game $c$ is \emph{concave} if the TU-game $-c$ is convex. Notice also that the maxmin procedure defined for TU-games with strategies becomes the minmax procedure. Then, during this section $\psi$ denotes the minmax procedure.

The second example described in the Introduction corresponds to the model of airport games with strategies and illustrates the fact that $\psi(X,C)$ is not an airport game in general. Next,  we provide a sufficient condition for the transmission of the non-emptiness of the core in airport games with strategies. This condition is easier to check than the one given in Theorem~\ref{thm:core} because we do not need to compute any core.
\begin{theorem}\label{thm:a1}
	Let $(X,C)\in SG(N)$ be an airport game with strategies. If there is some $i_N\in N$  such that $\psi(X,C)(S)\geq \psi(X,C)(N)$, for every $S\subset N$ with $i_N\in S$, then $\psi(X,C)$ is balanced.
\end{theorem}
\begin{proof}
	Recall that the minmax procedure defines the following TU-game:
	\begin{equation*}
	\psi(X,C)(S)=\min_{x_S\in \prod_{i\in  S}X_i}\max_{x_{N\setminus S}\in\prod_{i\in N\setminus S}X_i}C(x)(S),\text{ for every }S\subseteq N.
	\end{equation*}
	Let $i_N\in N$ be such that $\psi(X,C)(S)\geq \psi(X,C)(N)$, for every $S\subset N$ with $i_N\in S$. Take  $d_i=\min\{\psi(X,C)(S)\ :\ i\in S\}$, for every $i\in N$. Notice that $d_i\geq 0$, for every $i\in N$ and we have $d_{i_N}=\psi(X,C)(N)$ due to the choice of $i_N$. Moreover, $d_i\leq d_{i_N}$, for every $i\in N$. 
	
	Now, we consider the airport game defined by the ordered collection $(d_1,\ldots,d_{i_N})$. Taking into account that $d_i\leq \psi(X,C)(S)$ for every $i\in S$, then $d(S)=\max\{d_i:\ i\in S\}\leq \psi(X,C)(S)$. Besides, $d(N)=\psi(X,C)(N)$ and $d$ has a non-empty core because it is an airport game. Then, $Core(d)\subseteq Core(\psi(X,C))$ and we obtain that $\psi(X,C)$ has a non-empty core. \qed
\end{proof}

The condition in  Theorem~\ref{thm:a1} holds, for instance, if there is some player $i_N$ with $C(x)(i_N)=\max\{C(x)(j)\ :\ j\in N\}$, for every $x\in X$ (i. e. if $i_N$ is a player with the most costly type of planes for all strategy profiles). Next result proves it.
\begin{corollary}
	Let $(X,C)\in SG(N)$ be an airport game with strategies such that there is some player $i_N$ with $C(x)(i_N)=\max\{C(x)(j)\ :\ j\in N\}$, for every $x\in X$. Then, $\psi(X,C)$ is balanced.
\end{corollary}
\begin{proof}
	Let $i_N$ be such that $C(x)(i_N)=\max\{C(x)(j)\ :\ j\in N\}$, for every $x\in X$. We prove that $i_N$ satisfies the condition of Theorem~\ref{thm:a1}. 
	Notice that there is some $\tilde{x}$ such that 
	\begin{equation*}
	\psi(X,C)(N)=\min_{x\in X}C(x)(N)=C(\tilde{x})(N)=\max_{j\in N}C(\tilde{x})(j)=C(\tilde{x})(i_N).
	\end{equation*}
	We prove that $\psi(X,C)(S)\geq \psi(X,C)(N)$ for every $S\subset N$ with $i_N\in S$. Let $S\subset N$ with $i_N\in S$.  Then, there is some $\bar{x}\in X$ such that 
	\begin{equation*}
	\begin{array}{lcl}
	\psi(X,C)(S)&=&\min_{x_S\in \prod_{i\in  S}X_i}\max_{x_{N\setminus S}\in\prod_{i\in N\setminus S}X_i}C(x)(S)=C(\bar{x})(S)\\
	&=&C(\bar{x})(i_N)\geq \min_{x\in X}C(x)(N)=C(\tilde{x})(i_N)=\psi(X,C)(N).
	\end{array}
	\end{equation*}
	Then, applying Theorem~\ref{thm:a1} we get the result. \qed
\end{proof}
The condition in Theorem~\ref{thm:a1} is sufficient, but not necessary as Example~\ref{ex:suff} illustrates.
\begin{example}\label{ex:suff}
	Let us consider $N=\{1,2,3\}$, $X_1=\{U,D\}$, $X_2=\{L,R\}$, and $X_3=\{F\}$. The airport (cost) games associated with each strategy profile are:
	\[\begin{array}{|r|ccccccc|}
	\hline
	S & \{1\} & \{2\} & \{3\} & \{1,2\} & \{1,3\} & \{2,3\} & \{1,2,3\} \\ \hline
	{C(U,L,F)} &   1   &   8   &   2   &    8    &    2    &    8    &     8     \\
	{C(U,R,F)} &   2   &   9   &   5   &    9    &    5    &    9    &     9     \\
	{C(D,L,F)} &   5   &  10   &   7   &   10    &    7    &   10    &    10     \\ 
	{C(D,R,F)} &   6   &   7   &   9   &    7    &    9    &    9    &     9     \\ \hline
	\end{array}\]                                                
	The minmax procedure provides the following TU-game:
	\[\begin{array}{|r|ccccccc|}
	\hline
	S & \{1\} & \{2\} & \{3\} & \{1,2\} & \{1,3\} & \{2,3\} & \{1,2,3\} \\ \hline
	\psi(X,C) &   2   &   9   &   9   &    7    &    5    &    9    &     8     \\ \hline
	\end{array}
	\]
	$\psi(X,C)$  is balanced since, for instance, the allocation $(0,3,5)$ belongs to the core of $\psi(X,C)$. Nevertheless, there is no player satisfying the condition of Theorem~\ref{thm:a1}. 
\end{example}
We have concave games for every strategy profile in an airport game with strategies, $(X,C)$. Despite this fact, the resulting TU-game $\psi(X,C)$ is not concave in general, as we can see in Example~\ref{ex:suff} by taking coalitions $\{1,2,3\}$, $\{1,2\}$, $\{2,3\}$ and $\{2\}$.
\subsection{Simple games with strategies}
 A \emph{simple game with strategies} is a pair $(X,V)\in SG(N)$ where $V(x)$ is a simple game for all $x\in X$. It means that for all $x\in X$ and  $ S\subset N$,
\begin{itemize}
	\item [$i)$]$V(x)(S)\in \{0,1\}$, for every $S\subset N$,
	\item [$ii)$] $V(x)(N)=1$, and
	\item [$iii)$] $V(x)(S)\leq V(x)(T)$, for every $S\subset T\subset N$.
\end{itemize}
Condition $iii)$ indicates that for every $x\in X$, $V(x)$ is a monotone game. As  a consequence of Theorem~\ref{thm:mon}, the TU-game built using the maxmin procedure, $\psi(X,V)$, is monotone. Besides, it is clear that $\psi(X,V)(N)=1$. Then, the maxmin procedure provides a simple game too. The third situation described in the Introduction is a particular case of a simple game with strategies. 

In general, the core of a simple game $w$ is empty unless  there is some veto player (i.e., a player $i\in N$ with   $w(N\setminus \{i\})=0$); besides, any allocation in the core of a simple game splits $w(N)$ among veto players. We  characterize the balancedness of the simple game resulting from the maxmin procedure applied to a simple game with strategies in the following result.
\begin{theorem}
	Let  $(X,V)\in SG(N)$ be a simple game with strategies. Then, the core of $\psi(X,V)$ is non-empty if and only if there is some $i\in N$ such that, for every $x_{N\setminus\{i\}}$,  player $i$ is a veto player in $V(x_{N\setminus \{i\}},\bar{x}_i)$ for some $\bar{x}_i\in X_i$.
\end{theorem}
\begin{proof}
First we assume that $Core(\psi(X,V))$ is non-empty. Then, there is a veto player $i$ in $\psi(X,V)$ because this TU-game is a simple game. Using Theorem~\ref{thm:core}, we have $Core(\psi(X,V))=\cap_{x\in X}Core(V^x)$ with 
	\begin{equation*}
	V^x(S)=\min_{\tilde{x}_{N\setminus S}\in \prod_{j\in N\setminus S}X_j}V(x_S,\tilde{x}_{N\setminus S})(S),
	\end{equation*}
	for every $S\subsetneq N$ and $V^x(N)=1$, for every $x\in X$.  Since $\emptyset\neq Core(\psi(X,V))=\cap_{x\in X}Core(V^x)$, we have $Core(V^x)\neq \emptyset$, for every $x\in X$. Besides, $V^x$ is a simple game, for every $x\in X$. Then, due to the choice of $i$, we have that  $i$ is a veto player in $V^x$,  for every $x\in X$. This means that   for every $x_{N\setminus \{i\}}$ there is some $\bar{x}_i\in X_i$ such that $V(x_{N\setminus \{i\}},\bar{x}_i)(N\setminus \{i\})=0$. This proves the condition.
	
Now we assume that there is some $i\in N$ such that, for every $x_{N\setminus \{i\}}$,  player $i$ is a veto player in $V(x_{N\setminus \{i\}},\bar{x}_i)$ for some $\bar{x}_i\in X_i$. Then, $Core(V^x)\neq \emptyset$, for every $x\in X$. Moreover, the allocation $a\in \mathbb{R}^n$ with $a_i=1$ and $a_j=0$ for every $j\in N\setminus \{i\}$ belongs to $Core(V^x)$, for every $x\in X$. Then, $\cap_{x\in X}Core(V^x)\neq\emptyset$ and using Theorem~\ref{thm:core},  we obtain that $Core(\psi(X,V))\neq\emptyset$. \qed
\end{proof}

\section{Concluding remarks}

In this paper we study situations where the players cooperate and, previously, can choose some strategy that changes the values of cooperation. In order to allocate the benefits of cooperation using this strategic power, we apply the maxmin procedure to obtain a new TU-game. This analysis generalizes the models that appear in \cite{Carpente05} in the sense that we consider a non-additive characteristic function  for each strategy profile. We characterize the procedure using adaptations of some well-known and intuitive properties. Additionally, we study under which conditions some properties of the initial TU-games are transmitted to the TU-game obtained as a result of applying the maxmin procedure. Finally, we provide some sufficient conditions to obtain a balanced TU-game in case of airport games with strategies and characterize the non-emptiness of the core in  case of simple games with strategies.

Other procedures to transform a TU-game with strategies into a TU-game could be introduced and compared with the maxmin procedure. One possibility is to consider the following optimistic version of the maxmin procedure.
		The {\em maxmax procedure} is the map $\bar{\psi}:SG\rightarrow G$ given by
		$\bar{\psi} (X,V)(S)=\max_{x\in X}V(x)(S)$
		for all $(X,V)\in SG(N)$ and all $S\subset N$.
Neither the superaditivity nor the balancedness property are transmitted by the maxmax procedure as we see in the following example. Let us take	$N=\{1,2\},\ X_1=\{A,B,C,D\}\ \text{and } X_2=\{E\}$. The TU-games associated with each strategy profile are:
\begin{equation*}
		\begin{array}{|r|ccc|} 
		\hline
		S & \{1\} & \{2\} & \{1,2\}\\
		\hline
		V(A,E) & 4& 3& 10\\
		V(B,E) & 2& 5& 9\\
		V(C,E) & 2& 4& 7\\
		V(D,E) & 6& 1& 7\\ \hline
		\end{array}
\end{equation*}
It is easy to check that $V(x)$ is superadditive  and balanced for every $x\in X$. Nevertheless, the maxmax procedure provides the TU-game: $v(1)=6$, $v(2)=5$, and $v(12)=10$, which is  neither superadditive nor balanced.

\begin{acknowledgements}
This work has been supported by the  Ministerio de Econom\'{i}a y Competitividad through  grants MTM2017-87197-C3-1-P, MTM2017-87197-C3-2-P, MTM2014-53395-C3-1-P, MTM2014-53395-C3-3-P, MTM2014-54199-P, and by the Xunta de Galicia through the European Regional Development Fund  (Grupos de Referencia Competitiva ED431C-2016-015 and   ED431C-2016-040, and Centro Singular de Investigaci\'on de Galicia ED431G/01).
\end{acknowledgements}

\bibliographystyle{spbasic2}      
\bibliography{tugamestrategies}   

\section*{Appendix}
\textit{Proof of Theorem~\ref{thm:proc}.}\\
		First, we show that $\psi$ satisfies the five properties. 
	Let $(X,V)\in SG(N)$ and $i\in N$ be such that $V(x)(i)=c$, for all 
	$x\in X$. Then it is clear that $\psi (X,V)(i)=c$, which shows that 
	$\psi$ satisfies individual objectivity.
	
	Now, take $(X,V),(X,\bar{V})\in SG(N)$ such that $V(x)\leq \bar{V}(x)$ for all $x\in X$. Then, it is clear that $\psi (X,V)\leq \psi (X,\bar{V})$. This proves that $\psi$ satisfies monotonicity.
	
	To see that $\psi$ satisfies irrelevance of weakly dominated strategies notice that, if strategy $x_i$ for player $i$ is weakly dominated in $S$ ($S\subset N$ with $i\in S$) for $(X,V)\in SG(N)$, then 
	
	\begin{eqnarray*}
		\psi (X,V)(S)&=&\max_{x_S\in\prod_{j\in S}X_j}\min_{x_{N\setminus S}\in\prod_{j\in N\setminus S}X_j}V(x_S,x_{N\setminus S})(S)\\
		&=&\max_{x_S\in\prod_{j\in S\setminus\{ i\}}X_j\times X_i\setminus\{ x_i\}}\min_{x_{N\setminus S}\in\prod_{j\in N\setminus S}X_j}V(x_S,x_{N\setminus S})(S)\\
		&=&\psi (X^{-x_i},V)(S)
	\end{eqnarray*}
	
	To check that $\psi$ satisfies irrelevance of weakly dominated threats, notice that 
	if strategy $x_j\in X_j$ of a player $j$ 
	is a weakly dominated threat to player $i\neq j$ in $S$ ($S\subset N\setminus\{ j\}$ with $i\in S$) for 
	$(X,V)\in SG(N)$, then
	
	\begin{eqnarray*}
		\psi (X,V)(S)&=&\max_{x_S\in\prod_{k\in S}X_k}\min_{x_{N\setminus S}\in\prod_{k\in N\setminus S}X_k}V(x_S,x_{N\setminus S})(S)\\
		&=&\max_{x_S\in\prod_{k\in S}X_k}\min_{x_{N\setminus S}\in\prod_{k\in N\setminus (S\cup \{ j\})}X_k\times X_j\setminus\{ x_j\}}V(x_S,x_{N\setminus S})(S)\\
		&=&\psi (X^{-x_j},V)(S)
	\end{eqnarray*}
	
	Finally, it is clear that $\psi$ satisfies merge invariance.
	
	We now proceed to show that any procedure satisfying the five 
	properties must coincide with $\psi$.
	Let $\phi :SG\rightarrow G$ be a procedure satisfying 
	individual objectivity, monotonicity, irrelevance of 
	weakly dominated strategies, irrelevance of 
	weakly dominated threats, and merge invariance.
	Take a TU-game with strategies $(X,V)$ and fix a 
	non-empty coalition $S\subset N$. If $S=N$, then 
	merge invariance, 
	irrelevance of 
	weakly dominated strategies, and  
	individual objectivity
	clearly imply that 
	$$\phi (X,V)(N)=\max_{x\in X}V(x)(N)=\psi(X,V)(N).$$ 
	Assume now that $S\neq N$. We will 
	prove that $\phi (X,V)(S)=\psi(X,V)(S)$.
	
	Take the game with strategies $(X^S,V^S)$
	that is obtained from $(X,V)$ by 
	considering the coalition $S$ as a single player 
	$[S]$. Because $\phi$ satisfies merge invariance, 
	we know that $\phi (X,V)(S)=\phi (X^S,V^S)([S])$. Since $\psi$ also satisfies merge invariance, it is enough to prove that $\phi (X^S,V^S)([S])=\psi (X^S,V^S)([S])$. We know that 
	$$\psi (X^S,V^S)([S])=\max_{x_S\in\prod_{j\in S}X_j}\min_{x_{N\setminus S}\in\prod_{j\in N\setminus S}X_j}V(x_S,x_{N\setminus S})(S).$$ 
	In order to avoid a cumbersome notation, denote ${\cal V}^S=\psi (X^S,V^S)([S])$. Besides, choose $\bar{x}=(\bar{x}_i)_{i\in N}$ a strategy profile such that $V(\bar{x})(S)={\cal V}^S$.
	
	Let $(X^S,V^{S1})$ be the game with strategies that is obtained from 
	$(X^S,V^S)$ by bounding the utility of player 
	$[S]$ from above by ${\cal V}^S$, i.e.
	
	$$V^{S1}(x_S, x_{N\setminus S})([S])=\min\{ V^S(x_S, x_{N\setminus S})([S]),{\cal V}^S\}$$
	for all $x\in X$, and $V^{S1}(x_S, x_{N\setminus S})(T)=V^S(x_S, x_{N\setminus S})(T)$ for all $x\in X$ and all $T\neq [S]$.
	Since $\phi$ satisfies monotonicity, 
	we know that $\phi (X^S,V^S)\geq \phi (X^S,V^{S1})$.
	
	Now, notice that  every strategy $x_S\neq\bar{x}_S$ of player $[S]$ is weakly dominated by $\bar{x}_S$ in $[S]$ for $(X^S, V^{S1})$. In fact, 
	$$V^{S1}(x_S,\hat{x}_{N\setminus S})([S])\leq{\cal V}^S=\min_{\bar{x}_{N\setminus S}\in\prod_{j\in N\setminus S}X_j}V(\bar{x}_S,x_{N\setminus S})(S)\leq V(\bar{x}_S,\hat{x}_{N\setminus S})(S)$$
	and then
	$$V^{S1}(x_S,\hat{x}_{N\setminus S})([S])\leq V^{S1}(\bar{x}_S,\hat{x}_{N\setminus S})([S])$$
	for all $\hat{x}_{N\setminus S}\in\prod_{j\in N\setminus S}X_j$.
	
	Define $(X^{S2},V^{S2})$ to be the game with strategies that is obtained from $(X^S,V^{S1})$ by deleting all the strategies of player $[S]$ except strategy $\bar{x}_S$. Since $\phi$ satisfies irrelevance of weakly dominated strategies we know that $\phi (X^S,V^{S1})([S])=\phi (X^{S2},V^{S2})([S])$. Next, notice that for every $j\neq [S]$ 
	it holds that every strategy $x_j\in X_j\setminus 
	\bar{x}_j$ is a weakly dominated threat to player 
	$[S]$ in $[S]$ because
	$$V^{S2}(\bar{x}_S,\hat{x}_{N\setminus(S\cup\{ j\})},x_j)([S])=V^{S2}(\bar{x}_S,\hat{x}_{N\setminus(S\cup\{ j\})},\hat{x}_j)([S])={\cal V}^S.$$
	Define $(X^{S3},V^{S3})$ to be the game with strategies that is obtained from $(X^{S2},V^{S2})$ by deleting all strategies 
	$x_j\in X_j\setminus \bar{x}_j$ for 
	every player $j\in N^S\setminus \{ [S]\}$. Since $\phi$ satisfies irrelevance of weakly dominated threats we know that $\phi (X^{S2},V^{S2})([S])=\phi(X^{S3},V^{S3})([S])$.

	In the game with strategies $(X^{S3},V^{S3})$ every player $j$ has exactly one strategy: $\bar{x}_j$. Hence we can 
	use individual objectivity of $\phi$ to derive that
	$$\phi (X^{S3},V^{S3}) ([S])={\cal V}^S.$$
	
	Putting everything together, we see that 
	we proved that 
	$\phi (X,V)(S)=\phi (X^S,V^S)([S])
	\geq \phi (X^S,V^{S1})([S])=
	\phi (X^{S2},V^{S2})([S])=
	\phi (X^{S3},V^{S3})([S])={\cal V}^S=\psi (X,V)(S)$.
	
	To finish the proof of the theorem, we need to show that $\phi (X,V)(S)\leq \psi (X,V)(S)$.
	
	Consider again the game with strategies $(X^S,V^S)$. We define now a new game with strategies $(X^{S4},V^{S4})$
	by adding a strategy 
	$x^*_i\not\in X_i$ for each player $i\in N\setminus \{[S]\}$.
	The strategies $x^*_i$ are introduced as additional 
	threats to player $[S]$. 
	We add these strategies one by one. Without loss of generality, 
	we assume that $N\setminus S=\{ 1,2,\ldots ,|N\setminus S|\}$. 
	
	We first define the game with strategies $(X^{*S1},V^{*S1})$, by adding strategy $x^*_1$ 
	for player $1$ in the following way: $X_1^{*S1}=X_1^S\cup \{ x^*_1\}$,  $X_i^{*S1}=X_i^S$, for all $i\in N^S\setminus\{ 1\}$, $V^{*S1}(x)=V^{S}(x)$, for all $x\in X^S$, and
	$$V^{*S1}(x_S,x^*_1,x_{N\setminus (S\cup \{1\})})(T)=\min_{x_1\in X_1}V^{S}(x)(T)$$
	for all $x\in X$ and all $T\subset N^S$. Clearly,  $x^*_1$ is a weakly 	dominated threat to player $[S]$ in $[S]$ for $(X^{*S1},V^{*S1})$. 
	Because $\phi$ satisfies irrelevance 	of weakly dominated threats, it holds that $\phi (X^{*S1},V^{*S1})([S])=\phi(X^S,V^S)([S])$.
	
	Now, let $2\leq j\leq k$ and suppose that we have added a 	strategy $x^*_i$ for each player $i=1,2,\ldots j-1$ and defined the corresponding games with strategies $(X^{*Si},V^{*Si})$ so that in each game $(X^{*Si},V^{*Si})$ strategy $x^*_i$ is a
	weakly dominated threat to player $[S]$ in $\{[S]\}$ and thus $\phi (X^{*Si},V^{*Si})([S])=\phi(X^S,V^S)([S])$. To obtain the game with strategies
	$(X^{*Sj},V^{*Sj})$ we add a strategy $x^*_j$ for player $j$ in the following way: $X_j^{*Sj}=X_j^{*Sj-1}\cup \{ x^*_j\}$, $X_i^{*Sj}=X_i^{*Sj-1}$, for all $i\in N^S\setminus\{ j\}$, $V^{*Sj}(x)=V^{*Sj-1}(x)$, for all $x\in X^{*Sj-1}$, and
	$$V^{*Sj}(x_S,x_{\{1,\hdots ,j-1\}},x^*_j,x_{N\setminus (S\cup\{1,\hdots ,j-1\})})(T)=\min_{x_j\in X_j^{*Sj-1}}V^{*Sj-1}(x)(T)$$
	for all $x\in X^{*Sj-1}$ and all $T\subset N^S$.
	Again, 
	strategy $x^*_j$ is a weakly 
	dominated threat to player $[S]$ in $[S]$ for $(X^{*Sj},V^{*Sj})$. 
	Because $\phi$ satisfies irrelevance
	of weakly dominated threats, it holds that $\phi (X^{*Sj},V^{*Sj})([S])=\phi (X^{*Sj-1},V^{*Sj-1})([S])=\phi (X^S,V^S)([S])$.
	
	The game with strategies $(X^{S4},V^{S4})$ is the game with strategies $(X^{*Sj},V^{*Sj})$ which emerges from the procedure 
	described above after a strategy $x^*_i$ has been added for each player 
	$i\in N^S\setminus \{[S]\}$. 
	
	Notice that 
	\begin{equation}
	\label{VS4}
	V^{S4}(x_{[S]},x^*_{N\setminus\{[S]\}})([S])=
	\min_{x_{N\setminus S}\in X_{N\setminus S}}
	V^S (x_{[S]},x_{N\setminus S})([S])
	\leq {\cal V}^S
	\end{equation}
	for all $x_{[S]}\in X_{[S]}$. We will use this later.
	
	Let $(X^{S4},V^{S5})$ be the game that is obtained from 
	the game $(X^{S4},V^{S4})$ by bounding the utility of player 
	$[S]$ from below by ${\cal V}^S$, i.e.
	$$V^{S5}(x_S, x_{N\setminus S})([S])=\max\{ V^{S4}(x_S, x_{N\setminus S})([S]),{\cal V}^S\}$$
	for all $x\in X^{S4}$, and $V^{S5}(x_S, x_{N\setminus S})(T)=V^{S4}(x_S, x_{N\setminus S})(T)$ for all $x\in X^{S4}$ and all $T\neq [S]$.
	Since $\phi$ satisfies monotonicity, 
	we know that $\phi (X^{S4},V^{S5})\geq \phi(X^{S4},V^{S4})$.

	Now, notice that  every strategy 
	$x_i\in X_i$ is a weakly dominated threat to player 
	$[S]$ for every player $i\neq [S]$ in $(X^{S4},V^{S5})$ (it is weakly dominated by $x^*_i$).
	Because $\phi$ satisfies irrelevance
	of weakly dominated threats, it holds that $\phi (X^{S6},V^{S6})([S])=\phi (X^{S4},V^{S5})([S])$, where 
	$(X^{S6},V^{S6})$ is the game that is obtained from 
	$(X^{S4},V^{S5})$ by deleting all strategies 
	$x_i\in X_i$ for every player $i\in N^S\setminus \{[S]\}$.
	
	In the game $(X^{S6},V^{S6})$ all players $i\neq [S]$ have only 
	one strategy, strategy $x^*_i$, and since
	$V^{S6}(x_{[S]},x^*_{N\setminus\{[S]\})}([S])
	\leq {\cal V}^S$ (see (\ref{VS4})) and $V^{S6}\geq{\cal V}^S$, the  individual objectivity implies that $\phi(X^{S6},V^{S6})([S])={\cal V}^S$.
	
	Putting everything together, we see that 
	we proved that 
	$\phi(X,V)(S)=\phi (X^S,V^S)([S])=
	\phi (X^{S4},V^{S4})([S])\leq 
	\phi (X^{S4},V^{S5})([S])=
	\phi (X^{S6},V^{S6})([S])=
	{\cal V}^S=\psi (X^S,V^S)(S)$.
	
 Thus, we obtain the result.

\end{document}